# Collective behaviour of partons could be a source of energetic hadrons

M K  Suleymanov

COMSATS Institute of Information Technology, Park Road Chak Shahzad Town, Islamabad, Pakistan; Institute of Physics ANAS,  33, H.Javid  ave., Baku, Azerbaijan; Joint Institute for Nuclear Research, Dubna, Moscow region, Russia

E-mail: mais_suleymanov@comsats.edu.pk

**Abstract.** We discuss the idea that collective behaviour of the quarks/partons, which has been intensely discussed for the last 40 years in relativistic hadron-nuclear and nuclear-nuclear interactions and confirmed by new data coming from the ultrarelativistic heavy ion collisions, can lead to energetic particle production. Created from hadronization of the quark/parton (or quarks/partons), energetic particles could get the energy of grouped partons from coherent interactions. Therefore, we think that in the centre of some massive stars, a medium with high density, close to Quantum Chromodynamic one could be a source of the super high-energy cosmic rays.

## 1.Introduction
Cosmic rays are one of the main components of the Universe. They could provide crucial information about the beginning of the Universe and its evolution. The figure 1 demonstrates the variety of cosmic rays in a diagram displaying the flux of cosmic particles as a function of their energy (the figure was taken from the [1]). Even though, there have been a lot of research conveyed on cosmic rays and their applications, there is a lot about them that we don't know. One of these still unknown aspects is the source of cosmic rays with super high energies (especially in the range over $10^{15}$ $eV$) [2]; we don't even know for sure whether their origin is galactic or extragalactic. It was proposed that the massive stars could be the source of these rays and their super high energies could have resulted from further acceleration by very strong magnetic fields generated by these stars [3]. However, theoretical predictions show that these magnetic fields could be too weak to accelerate particles to energies of order $10^{15}$ $eV$. So it is natural that we look for an alternative source of these super high-energy cosmic rays, which does not involve acceleration [4]. The presented paper focuses on one of the possible alternative sources. We propose that these particles could be produced by same massive stars, however without any acceleration, as a result of a collective behavior of quarks/partons through the coherent interactions in the extreme (high density and/or high temperature) states (possibly from the Quark Gluon Plasma) of strongly interacting matter.

## 2. Phases of strongly interacting matter
It is considered, that during first $\sim 10^{-(22-23)}$ *sec* of nuclear-nuclear and hadron-nuclear interactions at relativistic and ultrarelativistic energies, the matter produced exists as a strongly interacting one, and this new system contains only hadrons. The figure 2 shows the Quantum Chromodynamics phase diagram for the strongly interacting matter (the figure was taken from the [5]). Expectations are such, that with increase in density the nuclear medium will transform into extreme dense states of the strongly interacting matter, and then it will pass a mixed phase around critical density (here we expect the appearance of the quark-parton degree of freedom). After that we expect matter to be in the onset states of deconfinement [5], which could occur because of percolation effect [6].

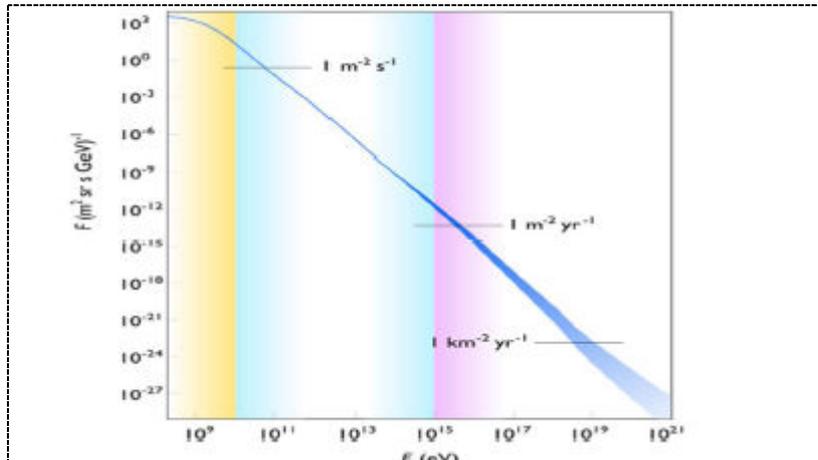

**Figure 1.** The flux of cosmic ray particles as a function of their energy. The flux for the lowest energies (yellow zone) are mainly attributed to solar cosmic rays, intermediate energies (blue) to galactic cosmic rays, and highest energies (purple) to extragalactic cosmic rays.

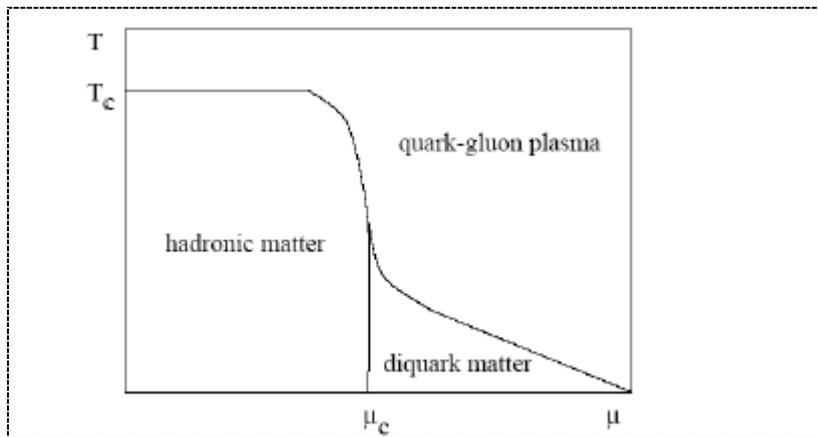

**Figure 2.** The phase diagram of strongly interacting matter.

### 3. Some evidence from the experiments.
The experimental data on relativistic and ultrarelativistic hadron-nuclear and nuclear-nuclear interactions have been showing the collective behavior for the quarks/partons in the hot and dence matter [7-12]. We believe that as a result of coherent interactions in the system, quark(s)/parton(s) can get the energy of the grouped quark/parton system, therefore the only limiting factor on final maximum energy of these quark(s)/parton(s) is the energy of the grouped quarks/partons in the system. The quarks/partons can hadronize and become super high-energy hadron cosmic rays.

The following effects we consider as some experimental evidence on quarks/partons collective phenomenon: *JINR* Cumulative effect [7] ; *CERN EMC* effect [8]; recent *RHIC* [9-11] and LHC [12] effects on saturation of $v_2$ as a function of the $p_T$.

*3.1. JINR Cumulative effect.*
When we say "energetic" (or "cumulative") hadrons we think about hadrons formed as a result of hadronozation of quark(s)/parton(s), that can get the grouped quark/parton energy in the high density (and/or temperature) medium. A.M.Baldin first predicted the formation of energetic hadrons through the cumulative effect [7]. In the experiment at JINR Dubna it was shown that at energies of several *GeV* the particle production in nuclear collisions is set up in the asymptotic regime. [7] This observation was described as the achievement of invariability of physical picture of the secondary particle production in nuclear fragmentation with increasing collision energy or achievement of so-called limiting fragmentation of nuclei. The figure 3 [13] shows the energy dependence (*E*) for the invariant cross section of cumulative protons (with momentum in the range 0,4 – 1,0 *GeV/c* and with angles: $160^0$ - $164^0$ in lab frame) produced at different targets by pions and protons (blacked points). The $\sigma_{in}$ is total cross section for inelastic $\pi p$ and $pp$ interactions.

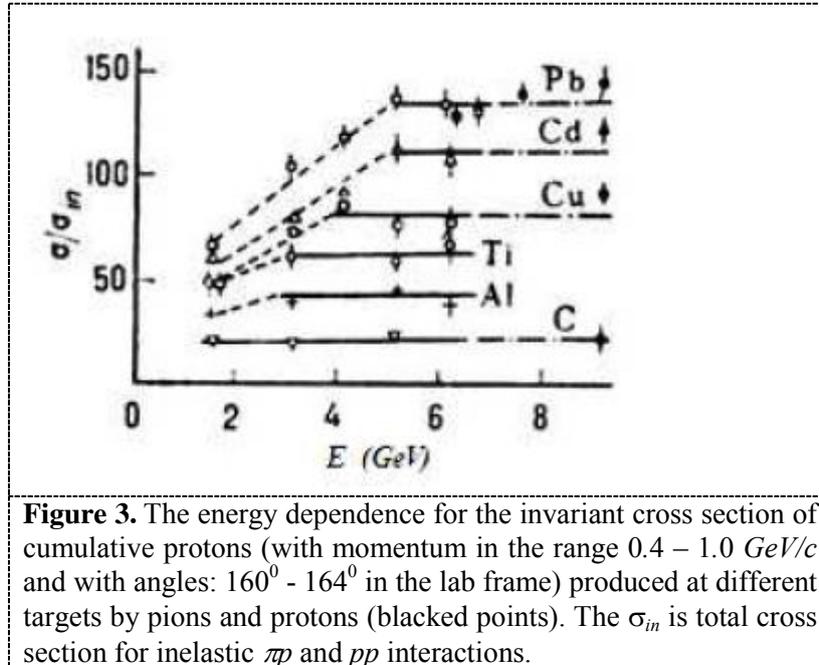

**Figure 3.** The energy dependence for the invariant cross section of cumulative protons (with momentum in the range 0.4 – 1.0 *GeV/c* and with angles: $160^0$ - $164^0$ in the lab frame) produced at different targets by pions and protons (blacked points). The $\sigma_{in}$ is total cross section for inelastic $\pi p$ and $pp$ interactions.

Cumulative processes involve production of secondary particles with energies beyond the kinematic limit of free nucleon collisions at relativistic hadron-nuclear and nuclear-nuclear interactions. This means that the secondary cumulative particles in these processes can be produced from systems of grouped nucleons, as a result of their collective behavior. Moreover, they get the total energies of the systems. In order to characterize these particles the cumulative parameter *x* was introduced, so that a secondary particles with *x > 1* are called cumulative ones. The following expression defines the value of *x*:

$$x = \frac{(E_i - P_L)}{m_N} , \quad (1)$$

where $E_i$, $P_L$ and $m_N$ are the energy, longitudinal momentum of a particle and $m_N$ is the mass of the nucleon respectively.

The value of $x$ is always less than $1$ for particles produced from free nucleon collisions. The figures 4 [14] and 5[15] show the $x$ and momentum spectra, respectively, for secondary charged particles produced in the hadron-nuclear reactions at relativistic energies. Appearance and existence of particles with $x>1$ can be observed from the figure 4. In figure 5 the arrow points on value of $x=1$.

Ref. [16] points to some theoretical explanation for the appearance of cumulative particles in hadron-nuclear and nuclear-nuclear interactions at relativistic energies. The authors propose that the effect is a result of nucleon collective phenomena.

The Coherent Tube Model (CTM) [17] can give us even a clearer explanation for the energetic

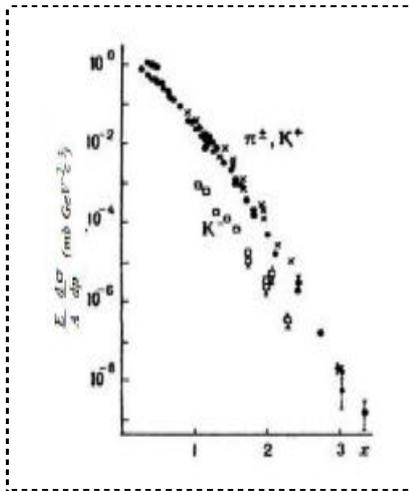
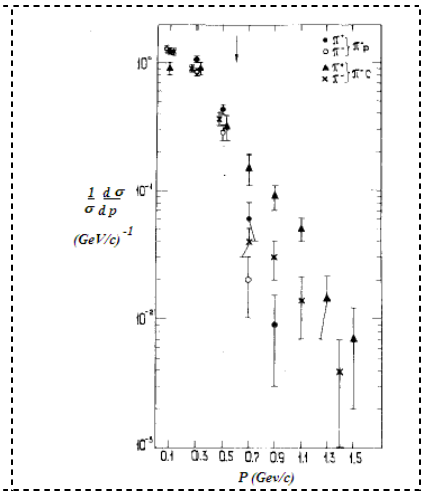

**Figure 4.** The $x$ spectra for secondary charged particles produced in the hadron-nuclear reactions at relativistic energies.

**Figure 5.** The momentum spectra for secondary charged particles produced in the hadron-nuclear reactions at relativistic energies.

(cumulative) particle production. Here the interaction of a hadron with a target nucleus results from its simultaneous collision with the tube of nucleons of cross section $\sigma$ that lie along its path in the target nucleus. For an interaction of projectile with momentum $p_{lab}$ the cumulative square of the center-of-mass energy is ~ $s_i \cong 2imp_{lab}$ ($i$ is a number of nucleons, $m$ - a nucleon mass). In paper [18] unusually strong $A$ dependence (stronger than commonly assumed $A$ or $A^{2/3}$) of the cross section for $p+ A - J/\Psi+X$ reaction at incident energies below $30\ GeV$ was described quantitatively using the cumulative effects (via energy rescaling).

Ref. [19] discusses CTM for high energy nucleus-nucleus collisions. In this case two tubes are considered: $i_1$ nucleons in incident tube and $i_2$ nucleons in the target tube. The c.m. energy squared for this tube-tube collision is approximately given by ~ $s_{i1i2} \cong 2i_1i_2 mp_{lab}$.

*3.2. CERN EMC-effect*
About 25 years ago, the deep-inelastic muon-nucleus scattering [8] at the European Muon Collaboration (*EMC*) at *CERN* revealed that the structure function $F_2$ and, hence, quark and gluon distributionof a nucleon bound in a nucleus differ for nuclei with different masses. The figure 6 shows the ratios of structure functions for different nuclei [20]. The paper [20] discusses that none of the popular models explaining the *EMC* effect seem satisfactory and presents a new point of view on the effect as a simple relativistic phenomenon.

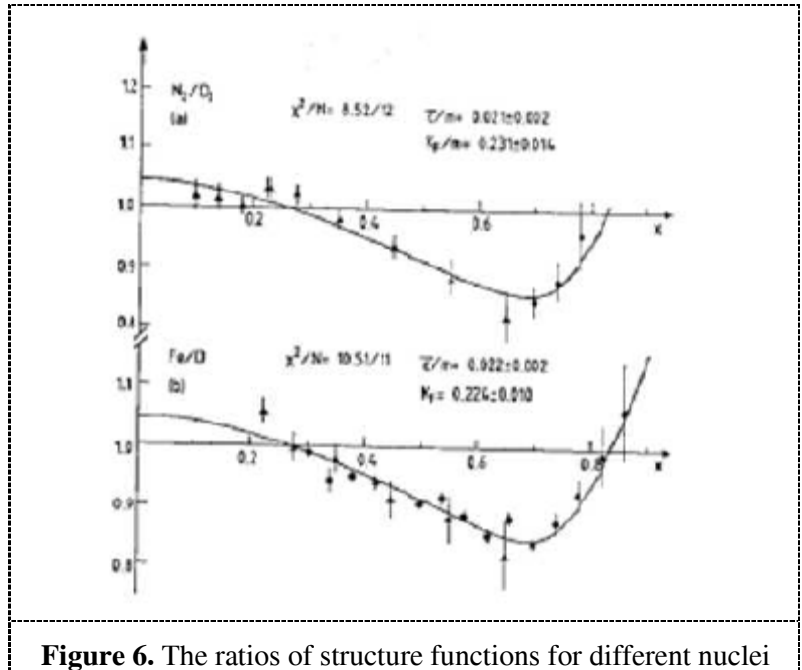
**Figure 6.** The ratios of structure functions for different nuclei

*3.3. Scaling behavior of $v_2$ at ultrarelativistic heavy ion collisions.*
Experimental results from studies of the azimuthal anisotropy of final states in nuclear-nuclear collisions at the energies of relativistic and ultrarelativistic heavy-ion colliders (RHIC, LHC) exhibit a collective behavior, which is likely to be formed at an early, parton, stage of the space-time evolution of produced hot and dense matter [9-11]. The data on azimuthal anisotropy indicates that strongly interacting matter produced in the final state under extreme conditions behaves as a nearly ideal liquid rather than ideal gas made of quarks and gluons.

Scaling behavior of $v_2$ (see figure 7 [10] and figure 8[11]) provides an opportunity to assume that the collective behavior of the partons defines the dynamics of an expansion in the longitudinal plane; evolution of the collective flow occurs on the parton state.

The first measurement of elliptic flow of charged particles in *Pb-Pb* collisions at the center of mass energy per nucleon pair $\sqrt{s}_{NN}$ = *2.76 TeV* (figure 8), with the *ALICE LHC* detector [12], shows that the $p_t$ –dependence of the $v_2$ does not change within uncertainties from *200 GeV* to *2.76 TeV*.

**4. Discussion and summary**
In the language of parton models, the facts lighted in previous chapters could point out that the high density (and/or temperature) nuclear matter can contain strongly correlated multiparton /multiquark formations and involve collective behaviour in the system.
As a result of coherent interactions quark(s)/parton(s) can get the energy of grouped quarks/partons (coherent nucleons or quarks/partons tube for example) and become a super high-energy quark(s)/parton(s). The quark(s)/parton(s) can leave the hot and/or dense medium with minimum energy lose and transform into a super high-energy hadron(s), where their energies are restricted only

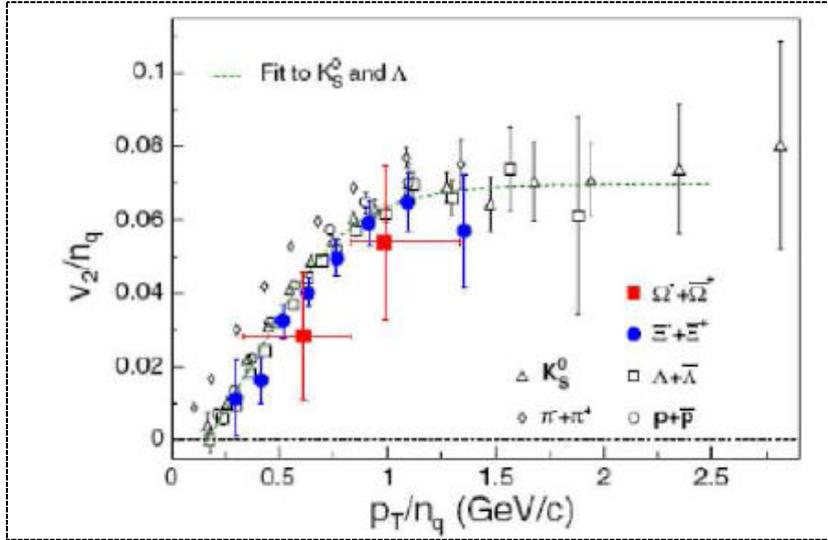

**Figure 7.** Number of quarks ($n_q$) scaled $v_2$ as a function of scaled $p_T$ for $\Xi+\Xi^+bar$ (filled circles) and $\Omega+\Omega^+bar$ (filled squares). Same distributions are also shown for $\pi\pi$ (open diamonds), $p+pbar$ (open triangles) [21], $K_S^0$ (open circles), $\Lambda+\Lambda bar$ (open squares) [22]. All data are from 200 $GeV$ $Au+Au$ minimum bias collisions. A dot-dashed-line is the scaled result of a fit to $K_S^0$ and $\Lambda$ [23].

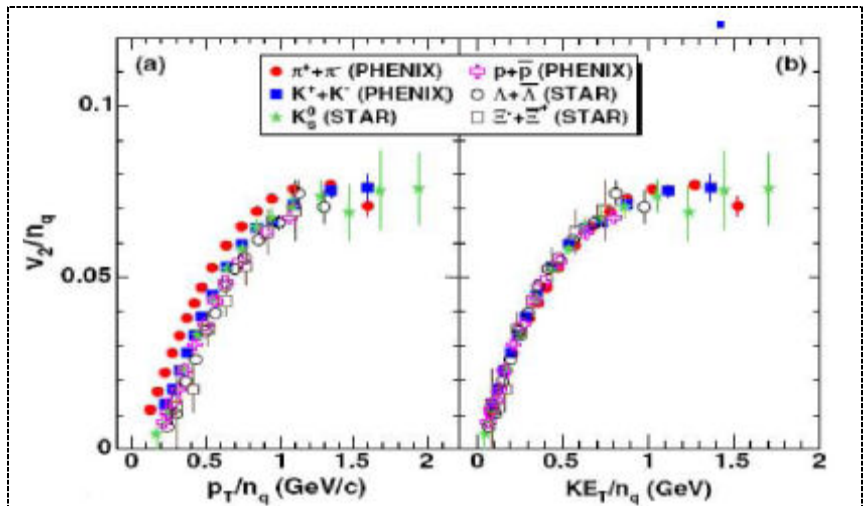

**Figure 8.** (a) $v_2=n_q$ vs $p_T=n_q$ and (b) $v_2=n_q$ vs $KE_T=n_q$ for identified particle species obtained in minimum bias $Au+Au$ collisions. The *STAR RHIC* data is from Refs. [24-25]

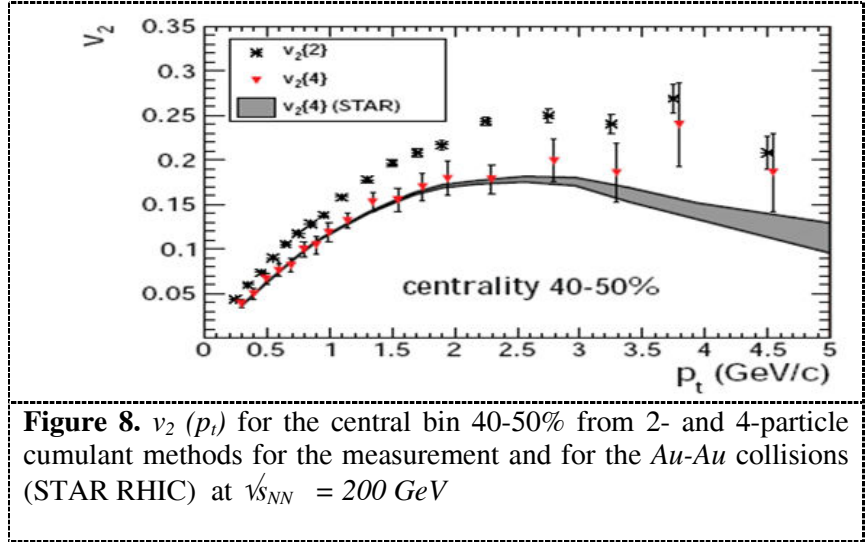

**Figure 8.** $v_2$ $(p_t)$ for the central bin 40-50% from 2- and 4-particle cumulant methods for the measurement and for the *Au-Au* collisions (STAR RHIC) at $\sqrt{s_{NN}} = 200\ GeV$

by the total energy of grouped quarks/partons system.

Now if we turn to cosmic rays, we can see that, it is possible for some medium with high density (and/ or temperature) close to the *QCD* critical one to be a source of cosmic particles with super high energies. This medium can evolve in the center of some massive star (e.g. neutron star [26]). High density (and/or temperature) of this medium can lead to the deconfinement. Quark(s)/parton(s) with big values of $x$ or energy can be formed in this system as a result of collective phenomenon and coherent interactions, where they hadronize and appear as super high-energy cosmic particles. As we have mentioned above, here the maximum energy of cosmic particles is limited only by the values of the total energy of the grouped quarks/partons.

Final notes:
1. Experimental results on relativistic and ultrarelativistic hadron-nucler and nuclear-nuclear collisions point out the collective behavior of the quarks/partons ;
2. Energetic quarks/partons can be produced from coherent interactions of the grouped quarks/partons;
3. In the center of some massive starts where the conditions are similar to ones suitable for the deconfinement, the produced energetic quarks/partons can transform into super high-energy cosmic rays, where their energies are limited only by the values of total energy of the grouped quarks/partons.

**Acknowledgments**

Author would like to thank Dr. S. M. Junaid Zaidi (Rector of CIIT, Islamabad) for his support for author's participation in the ICPP2011. Also thanks to A.M. Suleymanzade for help in preparing the paper.